\documentclass[12pt]{article}
\usepackage{pstricks}
\usepackage{graphicx}
\usepackage{axodraw}
\usepackage{epic}
\usepackage{amssymb,amsmath}
\def\beq#1\eeq{\begin{equation}#1\end{equation}}    
\def\bea{\begin{eqnarray}}  
\def\eea{\end{eqnarray}}  
\def\bq{\begin{quote}}  
\def\eq{\end{quote}}  
\def\bi{\begin{itemize}}  
\def\ei{\end{itemize}}  
\def\be{\begin{enumerate}}  
\def\ee{\end{enumerate}}  
\def\bi{\begin{itemize}}
\def\ei{\end{itemize}} 
\def\bc{\begin{center}}
\def\ec{\end{center}}

\def\cl{{\cal L}}

\def\r2{\sqrt{2}}  
\def\ra{\rightarrow}
\def\bi{\begin{itemize}}  
\def\ei{\end{itemize}}

\def\nn{\nonumber \\}

\def\tv#1{\vrule height #1pt depth 5pt width 0pt}
\def\tvbas#1{\vrule height 0pt depth #1pt width 0pt}
\def\ds{\displaystyle}

\newsavebox{\cmoose}
\sbox{\cmoose}{%
\begin{picture}(0,0)
  \thicklines
  \put(-60,0){\circle{35}}
  \put(60,0){\circle{35}}
  \ArrowLine(-40,0)(40,0)
  \ArrowLine(-60,-20)(-80,-60)
  \ArrowLine(-40,-60)(-60,-20)
  \ArrowLine( 80,-60)(60,-20)
  \ArrowLine( 60,-20)( 40,-60)
\end{picture}}

\newsavebox{\site}
\sbox{\site}{%
\begin{picture}(0,0)
  \thicklines
  \put(60,0){\circle{35}}
  \ArrowLine(-40,0)(40,0)
  \ArrowLine( 80,-60)(60,-20)
  \ArrowLine( 60,-20)( 40,-60)
\end{picture}}

\begin{document}
\pagestyle{empty}
\setcounter{page}{0}
{\normalsize\sf
\rightline {hep-ph/0203033}
\rightline{IFT-02/06}
\rightline{Saclay t02/022}
\vskip 3mm
\rm\rightline{March 2002}
}

\vskip 1.0cm
\begin{center}
{\huge \bf Soft electroweak breaking from hard}\\
\vskip.1cm
{\huge \bf   supersymmetry  breaking}
\vspace*{1cm}
\end{center}
 \noindent
\vskip 0.5cm
\centerline
{\sc Adam  Falkowski ${}^{1}$,
Christophe  Grojean ${}^{2}$
 {\rm and} Stefan  Pokorski${}^{1}$}
\vskip 1cm
\centerline{\em ${}^{1}$ Institute of Theoretical Physics, Warsaw University}
\centerline{\em Ho\.za 69, 00-681 Warsaw, Poland}
\vskip.2cm
\centerline {\em  ${}^{2}$ Service de Physique Th\'eorique, CEA--Saclay}
\centerline {\em F--91191 Gif--sur--Yvette, France}
\vskip.3cm
\centerline{\tt  afalkows@fuw.edu.pl, grojean@spht.saclay.cea.fr, pokorski@fuw.edu.pl}
\vskip 1.5cm

\centerline{\bf Abstract}
\noindent
We present a class of four-dimensional models,
with a non-supersymmetric spectrum,
in which the radiative corrections to the Higgs mass are
not sensitive, at least at one-loop, to the UV completion of the theory.
At one loop, Yukawa interactions of the top quark contribute
to a {\it finite} and {\it negative}  Higgs squared mass which
triggers the electroweak symmetry breaking, as in softly broken
supersymmetric theories, while gauge interactions
lead to a logarithmic cutoff dependent correction that can remain
subdominant. Our construction relies on a {\it hard} supersymmetry breaking localized in the
theory space of deconstruction models and predicts, within a renormalizable setup,
analogous physics as
five-dimensional scenarios of Scherk--Schwarz supersymmetry breaking.
The electroweak symmetry breaking can be calculated  in terms of the deconstruction scale, replication number, top-quark mass and electroweak gauge couplings.  For $m_{\rm top} \sim 170$~Gev,
the Higgs mass varies from 158~GeV for $N=2$ to 178~GeV for $N=10$.
\vskip .3cm


\newpage

\setcounter{page}{1} \pagestyle{plain}

The weak scale is unlikely to be a fundamental
scale of physics. Its calculation in terms of more
fundamental scales is one of the central problems in particle physics.
The problem is aggravated by the fact that in the Standard Model (SM)
the Higgs field mass parameter  gets radiative corrections that grow
quadratically with the scale $\Lambda$ of new physics:
$m_h^2 \sim \Lambda^2$. Thus, any attempt to calculate
$m_h$ in terms of a more fundamental scale $\Lambda$
and to make it stable against radiative corrections needs a mechanism of
suppression of the quadratic dependence on $\Lambda$;
the higher the scale $\Lambda$ is, the stronger suppression is required.

An attractive solution to the stability aspect of the hierarchy problem is provided
by softly broken supersymmetry. Quantum corrections to the weak scale depend
quadratically on $M_{\rm SUSY}$ and only logarithmically on the cutoff scale
$\Lambda$: $\delta_{m^2}\sim M_{\rm SUSY}^2 \ln \Lambda$
(therefore $\Lambda$ can be taken as high as the Planck scale).
However, the generation of the weak scale $v_h \ll \Lambda$ is overshadowed by the
$\mu$-problem, {\it i.e.}, by the question why the supersymmetric parameter
$\mu$ is of the order of the weak scale. Also, some other aspects of softly broken
supersymmetric theories are sufficiently troublesome to justify the quest for
alternative routes of solving the hierarchy problem.

An important new element in attempts to solve the hierarchy problem is the idea of  large
(TeV$^{-1}$ size) extra dimensions~\cite{AN},
 realized by low scale string theories or at the level
of effective theories~\cite{ANMUQU}.
One possibility, which does not require supersymmetry,
is to consider  the Higgs boson to be a component of the gauge field propagating
in extra dimensions. The Higgs potential, by higher-dimensional gauge invariance,
does not depend on the cutoff scale  and is calculable in terms of the compactification scale
$M_c \sim 1/R$~\cite{HO}.
Another possibility~\cite{ANDIPOQU,BAHANO,AHHNSW} is to break supersymmetry
{\it via} the
Scherk--Schwarz mechanism~\cite{SS}.
The non-local character of this mechanism  ensures that at least
one-loop corrections to the Higgs mass are finite~\cite{BAHANO,DEQU}.
In the effective theory supersymmetry is broken in a hard-way and it is conceivable that
divergences re-appear at higher-loop level.
However, large extra dimensions and related low value of the cutoff scale
$\Lambda$ change qualitatively the hierarchy problem in the sense that
calculating $m_h$ in terms of $R$ and $\Lambda$ does not require as strong
suppression of quadratic divergences as for the canonical case with
$\Lambda=M_{\rm PL}$. From a phenomenological point of view,
with the cutoff scale close to the compactification scale, one-loop finiteness of the
leading corrections to the Higgs potential is sufficient for the cutoff dependence
to be very weak. This point of view is taken in the model of Ref.~\cite{BAHANO}.
The electroweak symmetry breaking (EWSB) is triggered by  the top/stop loops.
Although the gauge interactions contribute to a  quadratically divergent result~\cite{GHNINI},
the finiteness of the leading corrections still allows to make predictions
about the Higgs boson mass~\cite{BHN2}.
The obvious advantage of this scenario is that the full Higgs potential and
the superpartner masses  are calculable to a good precision in terms of one
dimensionful parameter --- the compactification radius, and  the soft breaking
masses are {\it not necessary}.

Recently, a new idea called {\it deconstruction} appeared~\cite{ARCOGE_dec,HIPOWA}
which allows to realize the physics of extra dimensions in a strictly four-dimensional set-up.
Soon after, the 4D analogue of the mechanism~\cite{HO} was
constructed~\cite{ARCOGE_ele} where the Higgs boson  mass
is protected from receiving divergent radiative corrections by the pseudo-Goldstone
mechanism~(see also~\cite{ChHiWa} for another deconstruction model
of electroweak symmetry breaking and~\cite{twist} for a deconstruction model where
the radiative corrections are highly suppressed as a result of the topological
nature of the supersymmetry breaking).
Although the deconstruction models yield no unambiguous predictions about the
fundamental scale, the low-scale unification~\cite{unification}
suggests that the fundamental scale could be much lower than the Planck scale.
Thus, similarly as in the large extra dimensions models,
less suppression of the quadratic divergence is required to alleviate the hierarchy problem.
In this paper we investigate the four-dimensional analogue of the
Scherk--Schwarz mechanism and take the model proposed by Barbieri, Hall
and Nomura~(BHN)~\cite{BAHANO}
as our reference point. We do not aim at constructing a complete
and phenomenologically viable model, which would give the Standard Model
as its low-energy approximation. We  rather aim  at analyzing the general situation,
when divergences in non-supersymmetric theories are considerably softened.


More precisely, we  start with ${\cal N}=1$ supersymmetric
models consisting of a chain of $N$ gauge groups
which communicate to each other through $N-1$ bifundamental
link-Higgs fields $\Phi_i$~\cite{CSGRKRTE}.
The matter and Higgs fields are also replicated and
represented by a set of $N$ chiral superfields transforming
in fundamental representation  of the corresponding gauge group.
When the link-Higgs fields acquire vacuum expectation values
(vev) the mass pattern of the gauge, Higgs and matter fields is
 similar as in the theories with extra dimensions. 
  We are mainly interested in the models in
which the low-energy spectrum (zero-modes of the mass matrix)
shows no sign of supersymmetry but still the radiative corrections to
the mass parameters are weakly dependent on the cutoff scale.

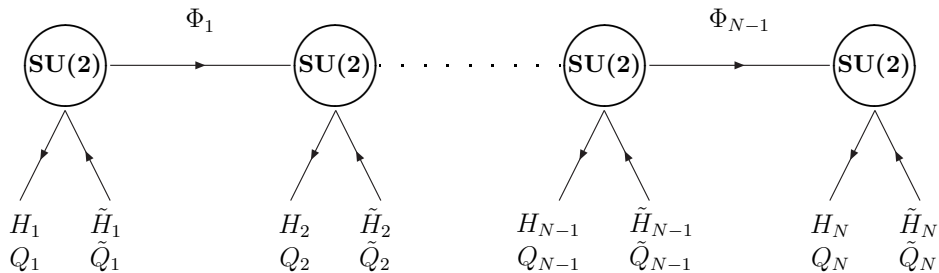
\begin{figure}[htb]
  \centering
  \scalebox{.85}{%
    \begin{picture}(-100,150)(0,0)
      \thicklines
      \put(-180,100){\usebox\cmoose}
      \put(60,100){{\usebox\cmoose}}
      \put(-60,100){{\makebox(0,0){\dottedline{10}(-40,0)(40,0)}}}
      \put(-240,100){\makebox(0,0){\small\bf SU(2)}}
      \put(-120,100){\makebox(0,0){\small\bf SU(2)}}
      \put(0   ,100)   {\makebox(0,0){\small\bf SU(2)}}
      \put( 120,100){\makebox(0,0){\small\bf SU(2)}}
      \put(-180,120){\makebox(0,0){\bf $\Phi_1$}}
      \put(  60,120){\makebox(0,0){\bf $\Phi_{N-1}$}}
       \put(-240,30){\makebox(0,0){\bf $H_1$\hspace{0.6cm} $\tilde{H}_1$}}
       \put(-240,15){\makebox(0,0){\bf $Q_1$\hspace{0.6cm} $\tilde{Q}_1$}}
       \put(-120,30){\makebox(0,0){\bf $H_2$\hspace{0.6cm} $\tilde{H}_2$}}
       \put(-120,15){\makebox(0,0){\bf $Q_2$\hspace{0.6cm} $\tilde{Q}_2$}}
   \put(0,30){\makebox(0,0){\bf $H_{N-1}$\hspace{0.6cm} $\tilde{H}_{N-1}$}}
   \put(0,15){\makebox(0,0){\bf $Q_{N-1}$\hspace{0.6cm} $\tilde{Q}_{N-1}$}}
   \put(120,30){\makebox(0,0){\bf $H_{N}$\hspace{0.6cm} $\tilde{H}_{N}$}}
   \put(120,15){\makebox(0,0){\bf $Q_{N}$\hspace{0.6cm} $\tilde{Q}_{N}$}}
    \end{picture}
  }
\caption{The quiver diagram of the model. Each circle (site) represents an
$SU(2)$ ${\cal N} =1$ Yang-Mills multiplet.
Each line pointing outwards a circle represents a chiral multiplet in the
fundamental representation of the group while a line pointing towards a circle
stands for a chiral multiplet in the anti-fundamental representation of the group.}
\label{fig:model}
\end{figure}

First, recall how the 5D  supersymmetric Yang-Mills theories on $S_1/Z_2$ are realized in the 4D
set-up~\cite{CSGRKRTE}. We write a supersymmetric Lagrangian for a chain of
$N$ gauge multiplets $(A_{i}^a,\chi_i^a)$
(with common gauge coupling $g_0$) and  $N$ chiral,
link-Higgs multiplets ($\Phi_i, \Psi_i$). In this paper all the gauge groups are $SU(2)$ and the diagonal subgroup is identified with the SM weak hypercharge group. The link fields are  
$2 \times 2$ complex matrices transforming in the
fundamental representation of the $i$-th and antifundamental
of the ($i$+1)-th gauge group .
The orbifolding procedure is accounted for by
the fact that the first and the last gauge groups are not linked,
thus the quiver diagram has 'topology' of the line segment (see Fig.~\ref{fig:model}).
The product group is spontaneously broken by the link-Higgs fields
which acquire a common expectation value
$\langle \Phi_i \rangle= v \mathbf{1}$. Diagonalizing the mass matrix, one finds
that the spectrum in the large $N$ limit is the same as in the 5D super-YM theory compactified on $S_1/Z_2$. In particular,  the link-Higgs degrees of freedom account
for completing ${\cal N}=1$ gauge multiplets up to ${\cal N}=2$ at
every massive level. 

To realize the SM matter and Higgs fields in the bulk we need to deconstruct 5D hypermultiplets. 
To this end, to every gauge group we attach a set of  chiral multiplets: 'Higgs doublets'
$H_i = (h_i, \psi_{Hi})$ and 'quark doublets'
$Q_i = (q_i, \psi_{Qi})$, in the fundamental of the $i$-th group and their mirror partners with opposite quantum numbers 
$\tilde{H}_i = (\tilde{h}_i, \tilde{\psi}_{Hi})$, $Q_i = (\tilde q_i, \tilde\psi_{Qi})$
which complete the spectrum to ${\cal N}=2$ hypermultiplets.
The superpotential is chosen as:
\beq
\label{eq:gs}
W =  \left(\sum_{i=1}^{N-1} y_i^h\tilde{H}_i\Phi_i H_{i+1}
- \sum_{i=1}^{N} m_i^h \tilde H_i H_i
+ \sum_{i=1}^{N-1} y^q_i\tilde{Q}_i\Phi_i Q_{i+1}
-  \sum_{i=1}^{N} m_i^q \tilde Q_i Q_i \right)
\eeq
 To complete the Standard Model
quark spectrum we need to add right-handed quark multiplets $U_i$ and $D_i$
and their mirrors. Since no color or hypercharge group is present in our toy-model
these fields are singlets. The superpotential is chosen as
$ W=  (\sum_{i=1}^{N-1} y_i^u\tilde U_i U_{i+1}
-  \sum_{i=1}^{N}m_i^u \tilde U_i U_i)$
and analogously for $D_i$.
In order that the hypermultiplet mass towers match that of the gauge multiplet one
has to fine-tune the parameters of the superpotential $y_i = g_0$, $m_i= g_0 v$.
For hypermultiplets implementing the $S_1/Z_2$ orbifold would  consist in removing the mirror multiplets $\tilde H_N$, $\tilde Q_N$, $\tilde U_N$, $\tilde D_N$ at the $N$-th site. In general, one removes those fields at the $N$-th site which in the 5D picture
have negative $Z_2$ parity.

In order to get the Yukawa interactions of the Higgs boson with the up-quarks
it is sufficient to add to the  superpotential the Yukawa term which involves
only the superfields from the first site (we omit in our notations the $SU(2)$
antisymmetric tensor used to build an invariant from two fundamental representations):
\beq
\label{eq:ys}
W = \lambda\, Q_1 H_1 U_1
\eeq
In the 5D picture this choice corresponds to {\it brane} Yukawa interactions at $x_5=0$. 
In the 4D set-up such a choice is stable in the large $N$ limit when ${\cal N}=2$ supersymmetry is recovered.
For finite $N$, when supersymmetry is broken, Yukawa interactions at other sites will be generated at higher-loop level.
At the moment we do not have the Yukawa interactions of the Higgs boson
with the down-quarks; we comment on this issue at the end of the paper.

At this point one could proceed towards the phenomenological models in the standard way,
that is  add soft terms to obtain splittings of the multiplets and to trigger the
electroweak symmetry breaking. In this paper we investigate an  alternative road.

As a first step we investigate  loop corrections to the Higgs boson mass from the
Yukawa couplings. Generically, the dominant contribution to the one-loop Higgs
mass is due to Yukawa interactions with the top quarks.
In SM this contribution is quadratically divergent, while in the MSSM the
quadratic divergence is canceled by the top squark loops. Here, we analyze
the set-up where such boson-fermion cancellation occurs  when  we break supersymmetry
{\it in a hard way} by removing some of the degrees of freedom in a non-supersymmetric way.
For the discussion of divergences it is irrelevant what is the precise pattern of the
breaking; the only important thing is  that the part of the Lagrangian
involving the fields of the  first site maintains the supersymmetric form.
In particular, we assume that all supertraces at the first site are vanishing.

\begin{figure}
\begin{center}
\begin{picture}(450,110)(-265,-45)
  \DashArrowLine(-265,10)(-229,10){3} \Vertex(-229,10){2}
  \Text(-255,20)[b]{$H$}
  \DashArrowArc(-200,10)(29,90,450){3}
  \Text(-200,51)[b]{$u_{(n)}, q_{(n)} $}
  \Text(-200,-23)[t]{$\tilde{q}_{(m)}, \tilde{u}_{(m)}$}
  \DashArrowLine(-171,10)(-135,10){3} \Vertex(-171,10){2}
  \Text(-145,20)[b]{$H$}
  \DashArrowLine(-105,10)(-69,10){3} \Vertex(-69,10){2}
  \Text(-95,20)[b]{$H$}
  \ArrowArc(-40,10)(29,90,450)
  \Text(-40,51)[b]{$\psi_{Q_{(n)}}$}
  \Text(-40,-23)[t]{$\psi_{U_{(m)}}$}
  \DashArrowLine(-11,10)(25,10){3} \Vertex(-11,10){2}
  \Text(15,20)[b]{$H$}
  \DashArrowLine(55,-19)(120,-19){3} \Text(65,-9)[b]{$H$}
  \DashArrowArc(120,10)(29,0,360){3} \Vertex(120,-19){2}
  \Text(120,47)[b]{$u_{(n)}, q_{(n)}$}
  \DashArrowLine(120,-19)(185,-19){3} \Text(175,-9)[b]{$H$}
\end{picture}
\caption{One-loop diagrams involving the top Yukawa coupling and
contributing to the squared mass of the Higgs boson.}
\label{fig:yukawa}
\end{center}
\end{figure}
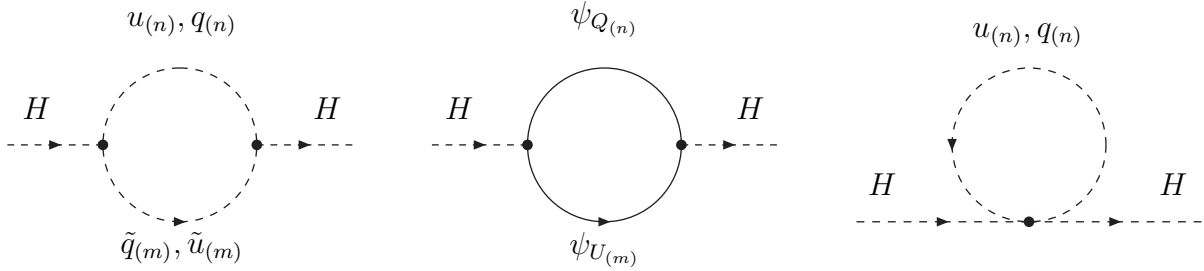

If the link-Higgs vevs are absent it is clear that at one-loop Yukawa interactions
do not feel the supersymmetry breaking on the other end of the chain.
Thus, the one-loop radiative correction to the $h_1$ squared mass
proportional to $\lambda$ are absent.
As soon as  we switch on the link-Higgs vevs, the fields living at different sites are allowed
to mix and we have to perform an orthogonal transformation to diagonalize the mass matrix.
Since supersymmetry is broken, generically the spectrum is completely  non-supersymmetric
(boson and fermion masses will be different and there can be a different number
of bosonic and fermionic degrees of freedom).
However, the $\lambda$-proportional corrections to the Higgs mass are still
controlled by  the first site
and, as a consequence, they are finite!
To see this we need to perform an orthogonal transformation  to express the original
fields in terms of the mass eigenstates:\footnote{From now, $x_i$ and $\psi_{X_i}$
will denote respectively the scalar and the chiral fermionic components
of the chiral superfield $X_i$. The mass eigenstates
will be denoted with parenthesized subscripts: $x_{(m)}$ and
$\psi_{X_{(m)}}$
with masses $m^x_{(m)}$ and $m^X_{(m)}$. The Lagrangians
involving fermions will be written in two component notation.  }
$q_i = \sum_n a^q_{i\, n} q_{(n)} $,  $\psi_{Q_i} =  \sum_n b^Q_{i\, n} \psi_{Q_{(n)}}$.
The zero mode Higgs mass receives one-loop radiative corrections
proportional to the Yukawa coupling through the diagrams depicted in Fig.~\ref{fig:yukawa},
which results in:
\bea
&&
-i\delta {m^2} = 2 \lambda^2  \int \frac{d^4k}{(2\pi)^4}
\left(
\sum_n  \frac{|a^q_{1\, n}|^2}{k^2 - {m^{q\, 2}_{(n)}}} +
\sum_{m,n} {m^q_1}^2  \frac{|a^q_{1\, n}|^2}{k^2 - {m^{q\, 2}_{(n)}}}
\frac{|a^q_{1\, m}|^2}{k^2 - {m^{q\, 2}_{(m)}}}
\vphantom{\frac{|b^Q_{1\,m}|^2}{k^2 - {m^{Q\, 2}_{(m)}}}}
\right.
\nn
&&
\hskip4cm
\left.
-k^2
\sum_{m,n} \frac{|b^Q_{1\,n}|^2}{k^2 - {m^{Q\,2}_{(n)}}}
\frac{|b^Q_{1\,m}|^2}{k^2 - {m^{Q\,2}_{(m)}}}
\right )
\eea

\noindent
The divergences in this expression are proportional to:
\bea
&&
\delta {m^2}
\sim
\Lambda^2 \
\left( \sum_n  |a_{1\, n}^q|^2 -\big(\sum_n  |b_{1n}^Q|^2\big)^2 \right)
\nn
&&
\hskip1cm
+\ln\Lambda^2 \
\left(
-  \sum_n  {m_{(m)}^{q\,2}} |a_{1\, n}^q|^2
- {m_1^q}^2 \big(\sum_n  |a_{1\,n}^q|^2\big)^2
+ 2 \sum_{m,n}
{m_{(m)}^{Q\,2}}  |b_{1\, n}^Q|^2 |b_{1\, m}^Q|^2
\right)
\eea

The coefficient of the quadratic divergence vanishes by the fact, that $a^q_{i\, n}$, $b^Q_{i\, n}$
are coefficients of the orthogonal transformation diagonalizing
the squark  and quark squared mass matrices, respectively. 
Furthermore, orthogonality identities from the diagonalization of the mass matrices
also give:
$\sum_n  {m^{q\,2}_{(n)}}  |a^q_{1\, n}|^2  =  \left({m^q}^2\right)_{11}$,
$\sum_n  {m^{Q\,2}_{(n)}} |b^q_{1\,n}|^2 = \left({m^Q {m^Q}^\dagger}\right)_{11}$.
This leads to the conclusion that the Higgs mass gets logarithmically
divergent contribution proportional to the supertrace in the quark sector
at the first site, which we assumed to vanish.
Thus, in spite of the fact that the theory is non-supersymmetric,
the Higgs mass (in fact the same holds for the squarks) gets, from the Yukawa interactions,
 only a finite one-loop correction to its mass. These conclusions hold even if the model has a different number of bosonic and fermionic degrees of freedom!

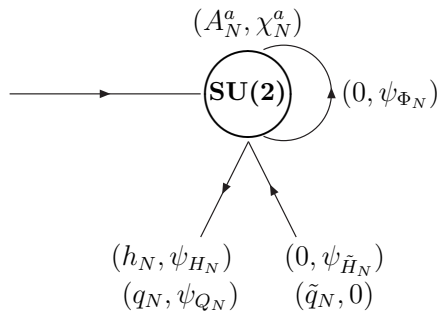
\begin{figure}[htb]
  \centering
  \scalebox{0.9}{%
   \begin{picture}(0,150)(0,0)
  \thicklines
  \put(-60,100){{\usebox\site}}
  \ArrowArc(15,100)(20,-115,115)
      \put( 0,100){\makebox(0,0){\small\bf SU(2)}}
      \put(60,100){\makebox(0,0){\bf $ (0,\psi_{\Phi_N})$}}
      \put(0,130){\makebox(0,0){\bf $ (A_N^a,\chi_N^a)$}}
   \put(0,30){\makebox(0,0){\bf $(h_{N},\psi_{H_N})$\hspace{0.6cm}
  $(0,\psi_{\tilde{H}_N}) $}}
   \put(0,15){\makebox(0,0){\bf $(q_{N},\psi_{Q_N})$\hspace{0.65cm} $(\tilde{q}_{N},0)$}}
    \end{picture}}
\caption[]{The magnifying glass view of the $N$-th site of the model.
The fields $\phi_N$, $\tilde{h}_N$ and $\psi_{\tilde{Q}_N}$ have been removed
in order to break supersymmetry and induce a mass splitting in the low-energy theory.
This specific construction  mimics the spectrum of the 5D supersymmetric model on
$S_1/Z_2 \times Z_2'$. Similarly as in the deconstruction of $S_1/Z_2$ models,
the orbifolding procedure amounts to removing those fields from
the mirror multiplets at the $N$-th site which in the 5D picture have both
$Z_2$ quantum numbers negative, the $(--)$ states of~\cite{BAHANO}.}
\label{fig:orbifolding}
\end{figure}

To illustrate this discussion we present  a construction inspired by the five-dimensional of
the BHN model~\cite{BAHANO}.
 Arriving at the spectrum of~\cite{BAHANO}  involves some tunings of the parameters.
But we stress that these tunings are by no means important for the
cancellation of divergences;
they serve only to obtain simple mass matrices,
so that formulae for the Higgs boson mass can be evaluated explicitly.
So we tune the parameters in the superpotential (\ref{eq:gs}) as:
\bea
y_i^h= y_i^q=y_i^d = y_i^u =g_0
\ \ {\rm and}\ \ \ m_i^h = m_i^q = m_i^d= m_i^u= g_0 v
\eea
We also add $\Phi_N$ link-Higgs, as in Fig. (\ref{fig:orbifolding}), which we need to avoid a massless gaugino mode.

In the second step we break supersymmetry by setting
$\phi_N$, ${\tilde h}_N$ and $\psi_{{\tilde Q}_ N}$ to zero
in the Lagrangian (see Fig.~{\ref{fig:orbifolding}}).
This is of course hard breaking of supersymmetry,
as some the fields at the last site lose their superpartners (a similar supersymmetry breaking
has also been proposed in Ref.~\cite{twist}).
Moreover, in order for the Higgs mass matrix to have a zero mode
we add a soft  breaking negative squared mass term, $-g_0^2 v^2 |h_N|^2$.\footnote{The last
step is not necessary  as the radiative corrections
will drive the lowest level higgs mass negative, even if  its tree-level
mass is not precisely zero.}
Diagonalizing the mass matrices yields a spectrum as in
Table~\ref{tab:spectrum}.
At the massless level we have only the gauge field, quarks and the Higgs boson.
Their lightest superpartners have masses
$\tilde m_{(1)}\sim g_0 v / (2N+1)$ and
include a Dirac gaugino, {\it two} squarks for every quark
and a Dirac Higgsino.
The complete mass matrices and their eigenvector can be found in~\ref{app:mode}.
Within this particular  model, we can now get a close expression of the finite
radiative Yukawa correction. Let us evaluate it.

\begin{table}
\large
\begin{center}
\begin{tabular}{|c|c|c|c|}
\hline
&gauge and link&quark&Higgs\\
\hline
$m_{(0)}=0$ & $A^a_{(0)}$ & $\psi_{Q_{(0)}}$ & $h_{(0)}$
\\
\hline
\tv{30}
\tvbas{20}
$m_{(n)} = 2 g_0 v \sin \left (\frac{n \pi}{2N} \right)$ &
 $A^a_{(n)}$, \,$\phi^a_{(n)}$ &
 $\left ( \begin{array}{c}\psi_{Q_{(n)}}\\
  -i \sigma_2\, \psi^*_{{\tilde Q}_{(n)}}\end{array} \right )$&
$h_{(n)}$,\,$ {\tilde h}_{(n)}$
\\
\hline
\tv{30}
\tvbas{20}
$\tilde{m}_{(n)} = 2 g_0 v \sin \left ( \frac{(2n - 1) \pi}{2(2N+1)} \right )$ &
 $\left ( \begin{array}{c} \psi_{\phi^a_{(n)}} \\
 -\sigma_2\, {\chi^{a\ *}_{(n)}}  \end{array} \right )$
 & $q_{(n)}$,\,$\tilde q_{(n)}$   &
$\left ( \begin{array}{c}\psi_{H_{(n)}}\\
-i \sigma_2\, \psi_{\tilde{H}_{(n)}}^*
\end{array} \right )$
\\\hline
\end{tabular}
\end{center}
\caption{{\small Spectrum of the model after the diagonal symmetry breaking.
The bifundamental link fields are decomposed into irreducible representations
of the diagonal $SU(2)$:
$\psi^a_{\phi_i}=\sqrt{2}\, \rm{tr}(T^a \psi_{\Phi_i})$ transforms
as an adjoint, while $\psi^s_{\phi_i}= \rm{tr}(\psi_{\Phi_i}) /\sqrt{2}$ transforms as a
singlet -- similar definitions for the bosonic components hold. Note that the singlet
bosonic as well as fermionic fields remain massless, fortunately they will also
remain uncharged under the full SM gauge group. }}
\label{tab:spectrum}
\end{table}

The top-quark Yukawa term (\ref{eq:ys}) in the superpotential  leads to the following
interactions:
\beq
\label{yukawa1}
\cl = - \lambda^2
|h_1|^2
(|u_1|^2+ |q_1|^2)
+ \lambda g_0 v
( h_1 u_1 \tilde{q}_1^\dagger  + h_1 q_1 \tilde{u}_1^\dagger + \rm{h.c.} )
-\lambda ( h_1 \psi_{U_1} \psi_{Q_1}+ \rm{h.c.})
\eeq

The orthogonal transformations into the mass eigenstates are given in   \ref{app:mode}.
Plugging in the mode decomposition we find the following interactions involving
the zero mode Higgs doublet $h_{(0)} \equiv H$:
\bea
\label{yukawa}
\cl =
 - \frac{4\lambda^2}{N(2N+1)} |H|^2
 \sum_{n,m=1}^{N}(u_{(n)} u_{(m)}^\dagger  + q_{(n)} q_{(m)}^\dagger )
  \cos\left( \frac{(2n-1)\pi}{4N+2}\right)  \cos\left( \frac{(2m-1)\pi}{4N+2}\right)
 \nn
 + \frac{4\lambda}{\sqrt{N}(2N+1)} \sum_{n,m=1}^{N}
( H u_{(n)}\tilde{q}_{(m)}^\dagger + H q_{(n)}\tilde{u}_{(m)}^\dagger + \rm{h.c.})
\tilde{m}_m  \cos\left( \frac{(2n-1)\pi}{4N+2}\right)  \cos\left( \frac{(2m-1)\pi}{4N+2}\right)
\nn
-\frac{2\lambda}{\sqrt{N}} \sum_{n,m=0}^{N-1}
(H \psi_{U_{(n)}} \psi_{Q_{(m)}} + \rm{h.c.}) \eta_n \eta_m
\cos\left (\frac{n \pi}{2N}\right) \cos\left(\frac{m \pi}{2N}\right)
\eea

\begin{figure}[ht]
\centerline{\epsfxsize=\textwidth\epsfbox{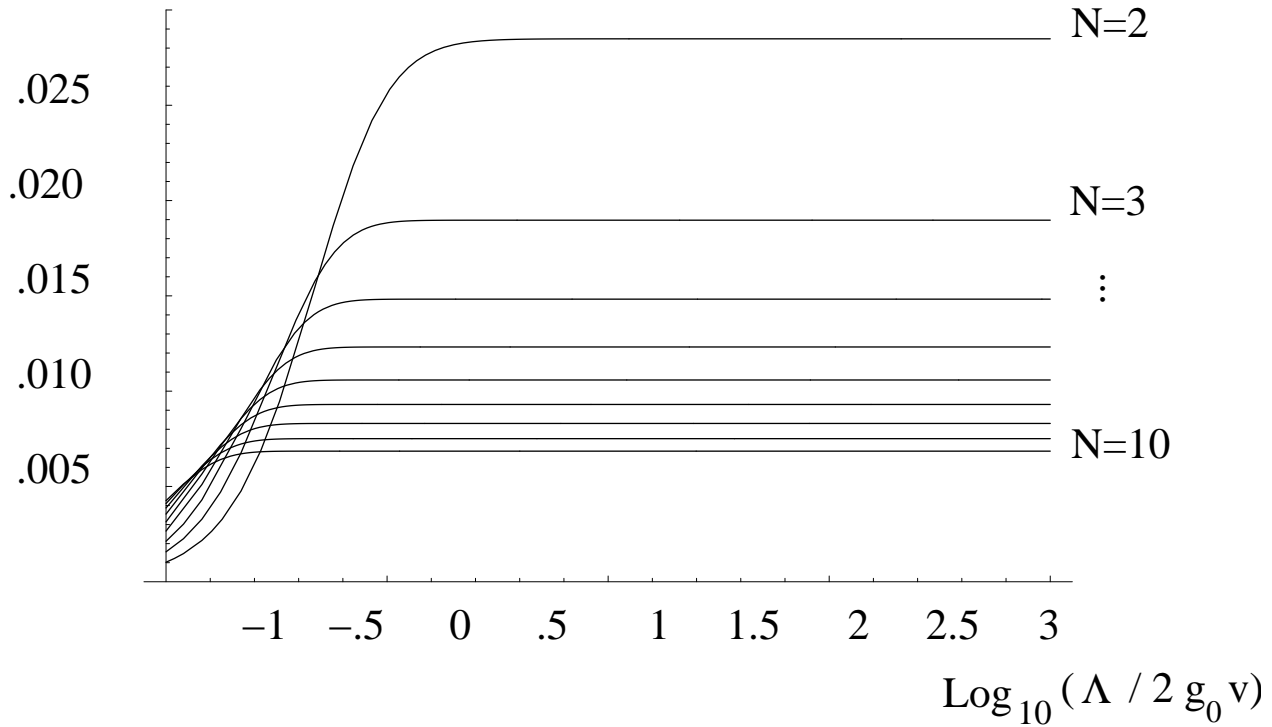}}
\caption[]{UV insensitivity of the one-loop Yukawa correction to the Higss mass.
We have plotted for different replication number, $N$, the numerical
factor that enters in the one-loop correction to the Higgs mass as a function of the cutoff
scale. We conclude that this factor is completely determined by the IR physics.
}
\label{fig:Yuk}
\end{figure}

The one-loop radiative correction to the Higgs mass can now be computed
explicitly and after some algebra, we arrive at:
\bea
\delta m^2 = -\lambda_T^2 (g v)^2 \, F(\Lambda,N)
\eea
where $F(\Lambda,N)$ is a pure numerical factor given by:
\bea
\label{eq:yukawamass}
F(\Lambda,N) = \pi^{-2} N^3 \int_0^{X}
dx \cosh x \sinh^3 x\, \frac{(\cosh 2x +1)( \cosh 2x + 2 \cosh 4Nx -1)}{\sinh^2 2Nx\,
\cosh^2 (2N+1)x}
\eea
the cutoff, $X$, being related to the cutoff scale of the theory by
$\Lambda=2 g_0 v \sinh X$. The normalized coupling, $\lambda_T=\lambda/N^{3/2}$, is
the Yukawa coupling of the zero mode Higgs to the zero mode quarks, {\it i.e.}, the
Yukawa coupling of the effective SM; similarly, $g = g_0/\sqrt{N}$ is the zero mode
$SU(2)$ gauge coupling. Notice that according to our general discussion, $F(\Lambda,N)$
is finite when $\Lambda$ goes to infinity. The Fig.~\ref{fig:Yuk} illustrates
the insensitivity of the Higgs mass to the high energy physics.

Note that in the 4D model, the sums over the KK modes are finite and
one can freely exchange the sum with the integral.
Thus, contrary to the 5D models, one does not  rely on the procedure of
KK-regularization, which is sometimes questioned~\cite{GHNI} or at least
requires a careful treatment of the symmetries of the theory~\cite{COPI}.
In a sense, the 4D model can be considered a regularization of the
construction~\cite{BAHANO},
which truncates the KK-tower at a finite value,
but still preserves the finiteness properties.

We have shown in general that one-loop corrections in certain non-supersymmetric
theories can be surprisingly softened. What about two- and higher-loop corrections?
The one-loop cancellation of quadratic divergences depends crucially on the
tree-level equality of the Yukawa and 4-scalar couplings of the Higgs field on the first site.
However, due to the mass splitting between quarks and squarks these couplings are
renormalized differently.\footnote{More precisely, at one-loop the infinite
part of renormalization is equal, thus one-loop running of both coupling is the same.
The difference is in the finite part of the renormalization.}
Thus we expect quadratic  divergences to reappear at the two-loop level.

We come back to the general analysis of the model defined by the superpotential
(\ref{eq:gs}).
Except for the top-Yukawa couplings there are other sources of quadratic divergences
which  are proportional to the gauge coupling or to the Yukawa couplings to the link-Higgs.
Following the discussion of the top-Yukawa contributions we analyze the general conditions to avoid any quadratic divergences. 
The first potential source originates from the couplings of the Higgs field to the gauge multiplet
(and to itself in the D-term scalar potential) and the relevant coupling leading
to quadratic divergences are:
\beq
\cl =\sum_i
\left(
g_0^2 \, h_i^\dagger T^a T^b h_i \,  A_{\mu\,i}^a A_{\mu\, i}^b
- \frac{1}{2} g_0^2\, \left(h_i^\dagger T^a h_i\right)^2
+\left(
i\sqrt2 g_0\,  h_i^\dagger T^a  \psi_{H_i}\chi_i^a
+ \rm{h.c.}
\right)
\right)
\eeq
The second source comes from the F-term of the superpotential
(\ref{eq:gs}) and the dangerous terms which yield quadratic divergences are:
\beq
\cl = -\sum_i
\left(
y_{i-1}^{h\,2} \, |h_i|^2 |\tilde{h}_{i-1}|^2
-y_{i-1}^{h\,2} \, h_i^\dagger \phi_{i-1}^\dagger \phi_{i-1} h_i
-\left(
y_{i-1}^h \psi{\tilde{H}_{i-1}} \psi_{\Phi_{i-1}} h_i + \rm{h.c.}
\right)
\right)
\eeq

Here, the situation is qualitatively different than in the case of top-Yukawa contribution, as interactions occur at all sites.
To avoid quadratic divergences proportional to $g_0$
we have to ensure that at every site the  full Higgs multiplet interacts
with the full gauge multiplet. In particular we need $N$ gauginos. 
 When the link-Higgs fields get vevs,  the gauginos get Dirac masses through bilinear
coupling to the $N-1$ link-Higgsinos (the $N$-th link-Higgsino does not couple to gauginos as the $N$-th link-Higgs is absent). To avoid  massless gauginos we need to add supersymmetry breaking terms.
This can be done either in a soft way --- by adding the Majorana mass terms for gauginos or
in a hard way  --  coupling gauginos and link-Higgsinos {\it via} Dirac mass term.
In our specific model we chose to add  the Dirac mass
$i\sqrt{2} g_0 v \,\chi_N^a\, \rm{tr}(T^a \psi_{\Phi_N})$,
where $T^a$ are the normalized hermitian generators
of the fundamental representation of $SU(2)$.
 This gives us, in the large $N$ limit, the same mass spectrum of gauginos  as in~\cite{BAHANO}.

Similarly, for  quadratic divergences proportional to $y_i$ to be absent,
we need  full link and mirror multiplets to be present at the $i$-th site.
Note that, since $y_N \equiv 0$,  adding or removing links-Higgs and mirror degrees of
freedom at
the $N$-th site has no consequence for the divergence of the Higgs mass.
We used this fact in our model and placed the hard supersymmetry breaking sector at the
$N$-th site.

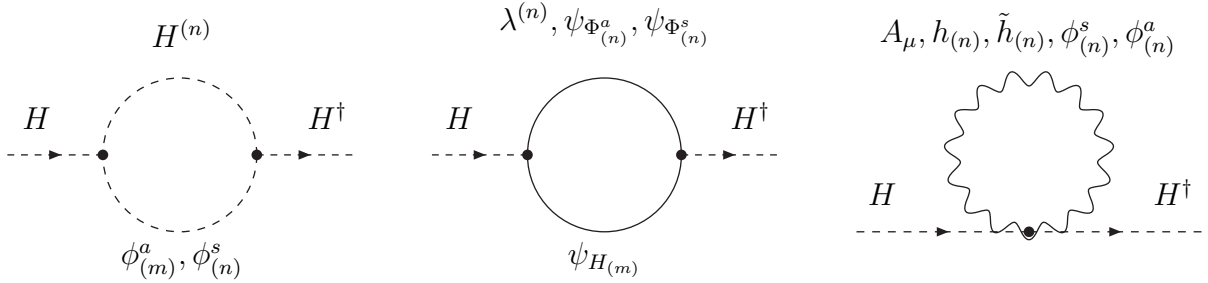
\begin{figure}
\begin{center}
\begin{picture}(450,110)(-265,-45)
  \DashArrowLine(-265,10)(-229,10){3} \Vertex(-229,10){2}
  \Text(-255,20)[b]{$H$}
  \DashCArc(-200,10)(29,0,360){3}
  \Text(-200,51)[b]{$H^{(n)}$}
  \Text(-200,-23)[t]{$\phi^a_{(m)},\phi^s_{(n)}$}
  \DashArrowLine(-171,10)(-135,10){3} \Vertex(-171,10){2}
  \Text(-145,20)[b]{$H^\dagger$}
  \DashArrowLine(-105,10)(-69,10){3} \Vertex(-69,10){2}
  \Text(-95,20)[b]{$H$}
  \CArc(-40,10)(29,0,360)
  \Text(-40,51)[b]{$\lambda^{(n)},\psi_{\Phi^a_{(n)}},\psi_{\Phi^s_{(n)}}$}
  \Text(-40,-23)[t]{$\psi_{H_{(m)}}$}
  \DashArrowLine(-11,10)(25,10){3} \Vertex(-11,10){2}
  \Text(15,20)[b]{$H^\dagger$}
  \DashArrowLine(55,-19)(120,-19){3} \Text(65,-9)[b]{$H$}
  \PhotonArc(120,10)(29,0,360){3}{15} \Vertex(120,-19){2}
  \Text(120,47)[b]{$A_\mu,h_{(n)},\tilde{h}_{(n)},\phi^s_{(n)},\phi^a_{(n)}$}
  \DashArrowLine(120,-19)(185,-19){3} \Text(175,-9)[b]{$H^\dagger$}
\end{picture}
\caption{One-loop diagrams involving gauge interactions and
contributing to the squared mass of the Higgs boson.}
\label{fig:gauge}
\end{center}
\end{figure}

In our specific model these interaction are all proportional to the gauge coupling
(as we tuned the link-Yukawa couplings with the gauge coupling).
The diagrams that contribute to the one-loop zero mode Higgs mass are depicted
in Fig~\ref{fig:gauge}. They give:
\bea
 \delta_{m^2} = -g^4 v^2 G(\Lambda,N)
 \eea
 where $g$ is the zero mode $SU(2)$ gauge coupling and
 $v$ is the deconstruction scale. The numerical factor $ G(\Lambda,N)$ is given
 in terms of a complicated integral:
\bea
 &&
 \hskip-1.1cm
 G(\Lambda,N) = \frac{N}{\pi^2}
 \int_0^{\Lambda/(2g_0 v)} dx \, x^3
 \left(
 \frac{7}{2} \sum_{n=1}^{N}
 \frac{1}{x^2 + \sin^2 \frac{(2n-1) \pi}{4N+2}}
 - \frac{7}{2} \sum_{n=1}^{N-1}
\frac{1}{x^2 + \sin^2 \frac{n \pi}{2N}}
-  \frac{3}{2} \frac{1}{x^2}
\right.
\nn
&&
\hskip-.9cm
\left.
- \frac{14}{2N+1}  \sum_{n=1}^{N}
 \frac{ \cos^2 \frac{(2n-1)\pi}{4N+2}}{x^2 + \sin^2 \frac{(2n-1) \pi}{4N+2}}
+ \frac{24}{(2N+1)^2}  \sum_{m,n=1}^{N}
\frac{x^2\, \cos^2 \frac{(2n-1)\pi}{4N+2}
 \cos^2 \frac{(2m-1)\pi}{4N+2}}{(x^2 + \sin^2 \frac{(2n-1) \pi}{4N+2})(x^2
 + \sin^2 \frac{(2m-1) \pi}{4N+2})}
\right)
\eea
As it can be checked explicitly and following our general discussion, this integral is
free of quadratic divergence. However, it exhibits a logarithmic divergence in the cutoff
scale:
\bea
\delta m^2 =
\frac{6}{8\pi^2} \, g^2 v^2 \, N \ln \left( \frac{\Lambda}{2 g v} \right)^2
\eea
The cutoff dependence is similar as in the softly broken supersymmetry,
but the $M_{\rm SUSY}$ scale is replaced here by the deconstruction scale $v$.
{  If $v$ is close to the weak scale (which is the case as long as $N$ is not too large)
then the one-loop sensitivity to the cutoff is weak.}


The previous evaluation of the Yukawa and gauge radiative corrections
to the zero mode Higgs mass
suggests that they will trigger the electroweak symmetry breaking. To study
in full details this breaking, we need now to compute the one-loop
effective potential given by:
\bea
V = \frac{1}{2} \, {\rm Tr} \, \int \frac{d^4 k_E}{(2\pi)^4}
\, \ln \frac{k_E^2+m_b^2 (v_H)}{k_E^2+m_f^2 (v_H)}
\eea
where $m_ b$ and $m_f$ are respectively the bosonic and fermionic
mass matrices as functions of the vev of the Higgs field, $v_H$.\footnote{In this paper,
our convention is to define $v_H$ as the vev of the complex Higgs field,
{\it i.e.}, $v_H\sim 174$\, GeV.}
We consider only the top-multiplet contribution and  the dependence on the Higgs vev is coming from the Yukawa interactions
localized on the first site only. 

For the {\it stop sector}, in the basis $(u_{(m)\,R}, \tilde{u}_{(n)\,L})$
and  $(u_{(m)\,L}, \tilde{u}_{(n)\,R})$, we obtain the following
$2N\times 2N$ squared mass matrix ($m,n,p,q=1\ldots N$):
\bea
        \label{eq:mb}
\left(
\begin{array}{cc}
\tilde{m}_{(m)}^2 \delta_{mp} + a_b\, c(m)\, c(p)
&
b_b\, \tilde{m}_{(m)}\, c(m)\, c(q)
\\
\\
b_b\, \tilde{m}_{(p)}\, c(m)\, c(p)
&
\tilde{m}_{(n)}^2  \delta_{nq}
\end{array}
\right)
\eea
where we have defined: $c(m)=\cos \frac{(2m-1)\pi}{4N+2}$, and the two coefficients
$a_b$ and $b_b$ are related to the Yukawa coupling as:
\bea
a_b = \frac{4\lambda^2 v_H^2}{(2N+1)N}
\ \ \  \mbox{and} \ \ \
b_b = - \frac{4\lambda v_H}{(2N+1)\sqrt{N}}
\eea
Note that squarks mix with mirror squarks once the electroweak symmetry is broken.

Similarly in the {\it top sector}, in the basis $(\psi_{\tilde{U}_L \, (m)}, \psi_{U_R\, (n)})$
and   $(\psi_{\tilde{U}_R \, (m)}, \psi_{U_L\, (n)})$, the
$(2N-1)\times (2N-1)$ squared mass matrix reads ($m,p=0\ldots N-1$,
$n,q=1\ldots N-1$):
\bea
        \label{eq:mf}
\left(
\begin{array}{cc}
m_{(m)}^2 \delta_{mp} + a_f\, \eta_m \eta_p \, d(m)\, d(p)
&
b_f \,\eta_m\, m_{(q)}\, d(m)\, d(q)
\\
\\
b_f \,\eta_p m_{(m)} \, d(m)\, d(p)
&
m_{(n)}^2  \delta_{nq}
\end{array}
\right)
\eea
where we have now defined $d(m)=\cos \frac{m\pi}{2N}$, and the two coefficients
$a_f$ and $b_f$ are given by:
\bea
a_b = \frac{2\lambda^2 v_H^2}{N^2}
\ \ \  \mbox{and} \ \ \
b_b = - \frac{2\lambda v_H}{N\sqrt{N}}
\eea

We will discuss in a moment the diagonalization of these matrices. But first, we would like to
show that two important supertraces are independent of the vev of the Higgs.
Indeed:
\bea
        \label{eq:STrM2}
{\rm STr_{\, top}}\, M^2
& = &
2 \left( \sum_{m=1}^N \tilde{m}_m^2 - \sum_{m=1}^{N-1} m_m^2 \right)
+ \frac{2N+1}{4} a_b - \frac{N}{2} a_f
 = 2\, g_0^2 v^2
 \\
{\rm STr_{\, top}} \, M^4
& = &
2 \left( \sum_{m=1}^N \tilde{m}_m^4   - \sum_{m=1}^{N-1} m_m^4 \right)
+ \frac{(2N+1)^2}{16} a_b^2
- \frac{N^2}{4} a_f^2
\nn
&&
\hskip1cm
+ 2 \left( a_b + \frac{2N+1}{4} b_b^2 \right)
\left( \sum_{m=1}^N \tilde{m}_m^2 \cos^2 \frac{(2m-1)\pi}{4N+2} \right)
\nn
&&
\hskip1cm
- 2  \left( a_f + \frac{N}{2} b_f^2 \right)
\left( \sum_{m=1}^N m_m^2 \cos^2 \frac{m\pi}{2N} \right)
\nn
        \label{eq:STrM4}
& = &
6\, g_0^4 v^4
\eea
This  ensures that the one-loop potential for
$v_H$ has no divergent dependence on the cutoff of the theory:
the EWSB is triggered  by the low energy physics and is not dominated
by unknown physics that will be revealed at or above the cutoff scale.
Adding the tree-level Higgs self-coupling originating from
the D-terms, we get:\footnote{Once the $U(1)$ hypercharge interactions
are included along the line discussed later, the gauge coupling, $g$,
in the tree level term  is  replaced by
$\sqrt{g^2+g'{}^2}$, where $g'$ is the $U(1)$ gauge coupling. The numerical simulations
reported at the end of the paper include of course the $U(1)$ interactions.}
\bea
V(v_H) = \frac{1}{8} g^2 v_H^4
+\frac{3}{16 \pi^2} \,
{\rm STr_{\,top}} \, m^4 \ln \left( \frac{m^2}{2g_0 v}\right)
\eea
where the supertrace is over the $2N$ bosonic and $2N-1$ fermionic
eigenvalues of the matrices~(\ref{eq:mb})--(\ref{eq:mf}).

Let us now turn to the determination of the spectrum. Using the general
formula (\ref{eq:det}) for the determinants, the secular
equation of the stop squared mass matrix is given by:
\bea
1 - \frac{16 \lambda^2 v_H^2}{N(2N+1)^2} \rho^2
\left( \sum_{m=1}^N \frac{\cos^2 \frac{(2m-1)\pi}{4N+2}}{\tilde{m}_{(m)}^2 - \rho^2}
\right)^2
=0
\eea
which, using some remarkable identities, can be rewritten as a polynomial
equation of degree $2N$:
\bea
RT_N (1-x^2) = \tau\, x\, RU_{N-1} (1-x^2)
\eea
where $x$ is the dimensionless eigenvalue $x=\rho/(2 g_0 v)$,
$\tau=\lambda_T v_H N / (g_0 v)$ characterizes the Higgs vev in units of~$g_0 v$,
and $RT_N$ and $RU_{N-1}$ are the reduced Chebyshev polynomials:
\bea
RT_N (X) = \frac{T_{2N+1} (\sqrt{X})}{\sqrt{X}}
\ \ \ \mbox{and} \ \ \
RU_{N-1} (X) = \frac{U_{2N-1} (\sqrt{X})}{\sqrt{X}}
\eea

Similarly, the fermionic secular equation is
\bea
1 - \frac{4 \lambda^2 v_H^2}{N^3} \rho^2
\left( \sum_{m=0}^{N-1} \eta_m^2\,
\frac{\cos^2 \frac{m\pi}{2N}}{m_{(m)}^2 - \rho^2}
\right)^2
=0
\eea
and it can be written in the form of
a polynomial equation of degree $2N-1$:
\bea
RT_{N-1} (1-x^2) = \tau^{-1}\, x\, RU_{N-1} (1-x^2)
\eea

\begin{figure}[ht]
\centerline{\epsfxsize=\textwidth\epsfbox{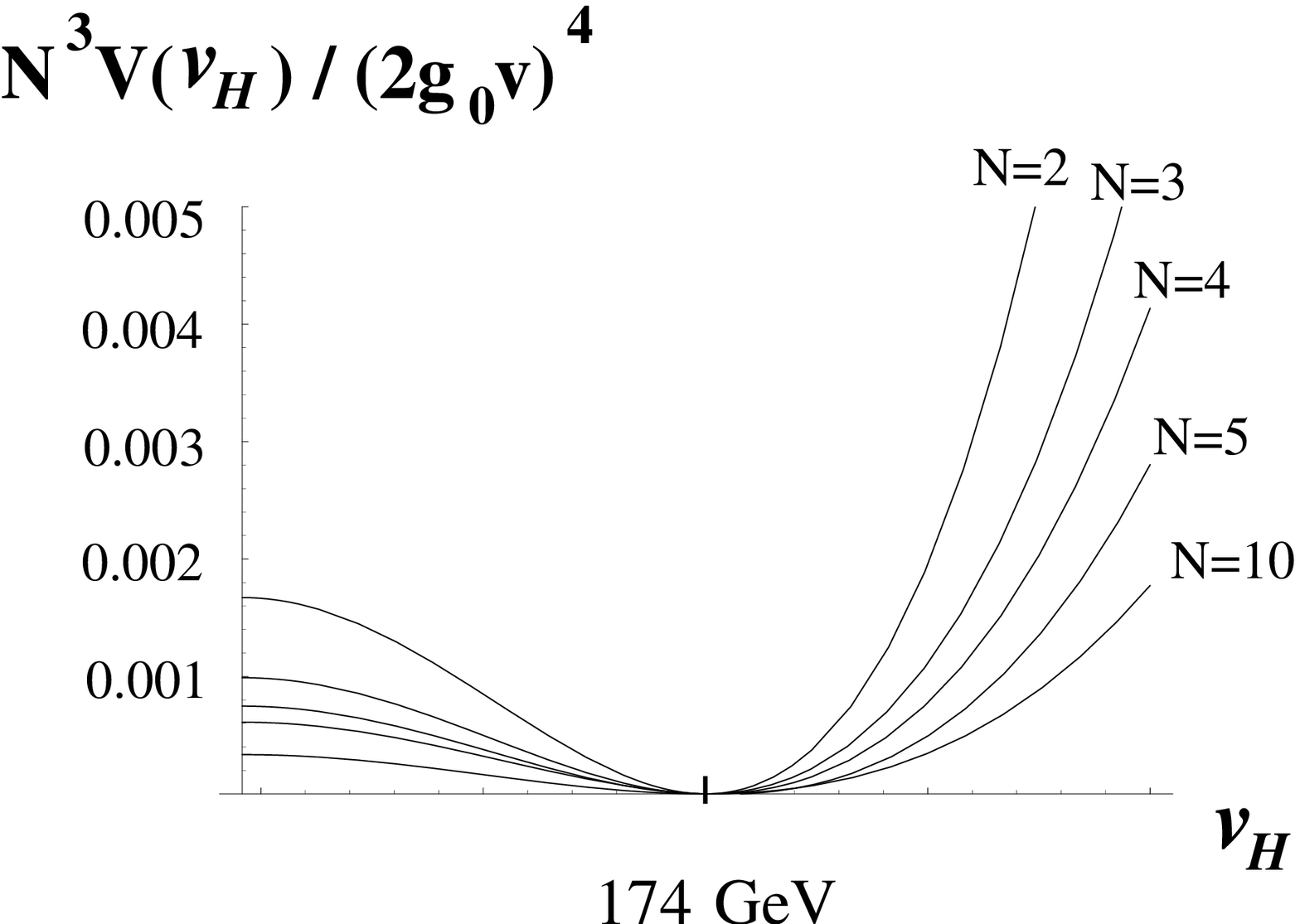}}
\caption[]{One-loop effective potential for the Higgs scalar field for different value
of the replication number $N$. As in softly broken supersymmetric theories,
radiative corrections trigger EWSB. The Higgs vev is a function of
the top Yukawa coupling, the SM gauge coupling constants and the deconstruction scale.
Fitting this Higgs vev to its phenomenological value, we get the prediction
for the Higgs mass. The UV insensitivity of the Higgs one-loop effective potential
ensures that the EWSB is reliably computed in our effective theory.
}
\label{fig:Pot}
\end{figure}

All the pieces are now set to numerically evaluate the potential $V(v_H)$ and find its
minimum. The results are plotted in Fig.~\ref{fig:Pot} for different values of the replication
number $N$.  The Higgs mass after EWSB becomes a function
of low energy parameters only: the top Yukawa coupling, the Higgs vev,
the electroweak gauge coupling and the replication number. The deconstruction
scale, $v$ is indeed fixed in terms of the low energy parameters once
the Higgs vev is fitted to its phenomenological value. Note that
the relation between the top mass and the top Yukawa coupling is modified
compare to the Standard Model since, here, the top mass corresponds to the lowest
eigenvalue of the fermionic mass matrix~(\ref{eq:mf}). Some numerical results
are given in Table~\ref{tab:results} for the Higss mass, the stop mass, the first
KK excitation of the top and the deconstruction scale, $v$. Interestingly enough,
after EWSB, the stop is lighter than the top.

\begin{table}
\begin{center}
\begin{tabular}{|c|c|c|c|c|}
\hline
& Higgs mass (GeV) & stop mass (Gev) & top first KK (Gev) & $v$ (GeV)
\\
\hline
$N=2$ & 158 & 142 & 502 & 437
\\
$N=3$ & 166 & 158 & 532 & 565
\\
$N=4$ & 170 & 161 & 533 & 664
\\
$N=5$ & 172 & 167 & 537 & 745
\\
$N=10$ & 178 & 164 & 517 & 1051
\\
\hline
\end{tabular}
\end{center}
\caption{{\small Spectrum of the model after the EWSB for different values
of the replication number, $N$. }}
\label{tab:results}
\end{table}

\vspace{.3cm}
Finally, we comment on how the toy-model presented here can be extended to
match the Standard Model phenomenology. The obvious missing pieces are:
\bi
\item
Leptons\\
One can  replicate the lepton superfields, $SU(2)$ doublets  $L_i$ and
singlets $E_i$,  and write superpotential exactly in the same way as for the quarks.
Analogously, one introduces Yukawa interactions of the leptons and the Higgs at the first site.

\item $SU(3)$ color group\\
In the real world quarks transform in the ${\bf 3}$ or ${\bf \bar 3}$
representation of the color group.
Replicating $SU(3)$ gauge group so as to obtain only one octet of massless gluon
and superpartners separated by a mass gap does not bring any complication.
One introduces a set of link-Higgs superfields $\Gamma_i$ transforming as
$({\bf 3_i,\bar 3_{i+1}})$ and the rest of the construction is analogous to the $SU(2)$ case.
The problem appears when we want to obtain the KK-tower for the quark doublets;
the gauge invariant superpotential must now have the form
$W = g_0\sum( \frac{1}{v} \tilde Q_i\Phi_i\Gamma_i Q_{i+1} - v\tilde Q_i Q_i)$
and leads to non-renormalizable interactions.
A more satisfactory alternative which allows to maintain renormalizability is
{\it not} to replicate the color gauge group and assume that all quark superfields
$Q_i$, $U_i$ and $D_i$ are charged under a single $SU(3)$.
Then the superpotential for these superfields has the same form as in the pure
$SU(2)$ case. Of course, then the model has no extra-dimensional interpretation
but this does not change any conclusions about softness of the radiative corrections.
It is nothing but a deconstructed version of a brane-world scenario where QCD
interactions are localized on the brane while weak interactions are free to propagate
in the bulk.

\item $U(1)$ hypercharge group\\ 
 Similar problems as in the $SU(3)$ case arise
when we replicate the hypercharge group:
writing an invariant superpotential so as to get the KK-tower of quarks and leptons
implies non-renormalizable interactions.
In addition, one must worry about anomalies, which must be compensated,
{\it e.g.}, by deconstructed Wess--Zumino terms~\cite{SKSM}.
Therefore, the more plausible alternative is not to replicate the hypercharge group.
One avoids non-renormalizable interactions and as a byproduct the anomalies
automatically cancel. Indeed, the fermion spectrum at all sites but the last one is vector-like:
every fermion is accompanied by the mirror fermion with opposite quantum numbers.
At the $N$-th site the fermion spectrum is the same as in the MSSM.
Note that in the 5D model of Ref.~\cite{BAHANO},
the issue of the $U(1)$ anomaly is more subtle~\cite{GHNINI,SCSESIZW,PIRI}.

\item Down-quark masses \\
In the supersymmetric models with only one light Higgs doublet there is the
well-known problem of coupling the neutral Higgs boson to the down quarks, as
holomorphicity forbids the $H^\dagger QD$ term in the superpotential.
A novel way to circumvent this problem was worked out in Ref.~\cite{BAHANO}.
It was noted that at the $Z_2'$ fixed point one can  write Yukawa interactions between
the multiplets of the {\it second} supersymmetry, which is not projected out at that point.
In the models of deconstruction this solution has no direct analogue, as the second
supersymmetry appears only in the large $N \ra \infty$ limit. Nevertheless, at the
$N$-th site one can still construct  multiplets
$H'_N = (h_N^\dagger,\psi_{\tilde{H}_N} )$,
$Q'_N = (\tilde q_N^\dagger,\psi_{Q_N})$,
$D'_N = (\tilde d_N^\dagger,\psi_{D_N})$
and write the Yukawa interactions {\it as if}
resulting from the superpotential term
$W =\lambda_D H'_N Q'_N D'_N$:
\bea
\label{yukawaN}
&
\cl = -
\lambda_D^2 \left(
\tilde{d}_N^\dagger \tilde{q}_N^\dagger \tilde{q}_N \tilde{d}_N
+ h_N^\dagger h_N \tilde{d}_N^\dagger \tilde{d}_N
+ \tilde{q}_N^\dagger h_N^\dagger h_N \tilde{q}_N
\right)
\nn&
\lambda_D g_0 v (
  h_N^\dagger \tilde q_{N}^\dagger {d}_{N} + h_N^\dagger  \tilde d_{N}^\dagger {q}_{N} + \tilde q_{N}^\dagger \tilde d_{N}^\dagger \tilde h_N+
{\rm h.c.})&\nn&
- \lambda_D
\left(
h_N \psi_{Q_N} \psi_{D_N}
+ \tilde{q}_N \psi_{\tilde{H}_N} \psi_{D_N}
+ \tilde{d}_N \psi_{\tilde{H}_N} \psi_{Q_N}
+\rm{h.c.}
\right)
\eea

It is easy to check that, similarly
as the Yukawa interactions (\ref{yukawa1})
at the first site, the quark and squark loops resulting
from (\ref{yukawaN}) do not produce any divergences in the Higgs mass.
\ei

So far supersymmetry is the best ingredient to protect the Higgs mass from
high energy physics. Unfortunately, a new sector plugged by unknowns and
even some fine-tunings is called for to break supersymmetry and to
allow for a phenomenologically viable spectrum.
Recently, Barbieri, Hall and Nomura proposed to
circumvent this supersymmetry breaking sector; the supersymmetry breaking would rather
have a geometrical
origin as resulting from a compactification on a non-trivial orbifold. We propose
a four-dimensional set-up, relying on a fully renormalizable theory, that shares similar
properties. Electroweak symmetry breaking is triggered by radiative corrections protected
by some remnants of supersymmetry, yet completely broken. The Higgs mass, at one loop,
is insensitive to UV physics and can be computed in terms of low energy parameters only.
In the past few years, several
non supersymmetric four-dimensional models have been
proposed~\cite{FRVA, Dienes,CrIbMa} in
which the one-loop Higgs potential is not destabilized by radiative
corrections, however higher loop corrections are much more dangerous if
not vexatious~\cite{CsSkTe}. Localization of supersymmetry breaking in the theory
space of deconstruction models might be here the key to protect the
predictability of the theory at higher loops.

\vspace{1cm}
{\bf Acknowledgments:}
The work of SP and AF was partially supported  by the EC Contract
HPRN-CT-2000-00148 for years 2000-2004. The work of SP was
partially supported  by the Polish State Committee
for Scientific Research grant KBN 5 P03B 119 20 for years 2001-2002.
SP and AF  acknowledge hospitality of  SPhT in Saclay where
part of this work was completed. CG thanks C.~Cs\'aki, J.~Erlich, G.~Kribs and
J.~Terning for fruitful discussions on similar constructions.

\renewcommand{\thesection}{Appendix~\Alph{section}}
\renewcommand{\theequation}{\Alph{section}.\arabic{equation}}
\setcounter{section}{0}
\setcounter{equation}{0}

\section{Mass matrices and mode decomposition}
\label{app:mode}

In this appendix we give the explicit formulae for the mass matrices and
mode decomposition into mass eigenstates in the model considered in the paper.
As it does not introduce any complication, we can be more general here and
consider a chain of $SU(K)$ groups with arbitrary $K$.

Let us begin  with the gauge sector.  We consider $N$ super-Yang-Mills $SU(K)$
multiplets ($A_i^a, \chi_i^a$) communicating to each other through
$N$ link-Higgs chiral multiplets ($\phi_i, \psi_{\phi_i}$) transforming as
$\bf (K_i, \bar{K}_{i+1})$; in other words links are $K\times K$
complex matrices transforming as $\Phi_i \ra U_i\Phi_i U_{i+1}^\dagger$.
The scalar $\phi_N$ is removed from the Lagrangian while the $N$-th link-Higgsino
transforms as $\bf (K_N, \bar{K}_{N})$. The D-term potential has vacua for
$\langle \Phi_i \rangle = v_i \mathbf{1}$.
If we take the special point on the moduli space where all
vevs are equal (and real) $v_i=v$, then the  product gauge group is broken down to
its diagonal subgroup. The mass term for the gauge bosons are~\cite{HIPOWA}:
\beq
\cl = \frac{1}{2} g_0^2 v^2 \sum_{i=1}^{N-1} (A^\mu_i - A^\mu_{i+1})^2
\eeq
This mass term is diagonalized by the following orthogonal transformation:
\beq
A_j =  \sqrt{\frac{2}{N}} \sum_{n=0}^{N-1}
\eta_n \cos \left (\frac{(2j-1) n \pi}{2 N} \right )
\, A_{(n)}
\eeq
  where $\eta_0=1/\sqrt2$ and otherwise $\eta_n=1$.
The spectrum of gauge boson contains a massless mode
$A_{\mu\, (0)}$ corresponding to the unbroken diagonal $SU(K)$ and a tower
of gauge bosons $A_{\mu\, (n)}$ with masses:
\beq
m_{(n)}  =  2 g_0 v \sin \left (\frac{n \pi}{2N} \right) \hspace{2cm} n=1 \dots N-1
\eeq
In the large $N$ limit the gauge boson masses are
$m_{(n)} \approx \frac{g_0 v \pi}{2 N} 2n$.
This matches the gauge boson spectrum of the BHN model upon identification:
\beq
\frac{1}{R} \sim \frac{g_0 v \pi}{2 N}
\eeq
This differs from the aliphatic case by a factor of one half,
which can be easily understood as resulting from orbifolding the circle twice.

The scalar link-Higgs degrees of freedom split into:
\bi
\item  $N-1$ real scalars  $G_i^a=iTr[T^a(\phi_i-\phi_i^\dagger)]$
in adjoint representation of the diagonal group.
These are massless (in the Landau gauge) Goldstone
boson which become the longitudinal components of the massive gauge fields.
Their counterpart in the 5D models are KK-modes of the fifth component of the gauge field.
\item  $N-1$ real scalars  $\phi_i^a=Tr[T^a(\phi_i +\phi_i^\dagger)]$
in adjoint representation of the diagonal group. When the link Higgs fields get vev,
the D-term potential generates mass terms:
\bea
\cl = \frac{1}{2}g_0^2 v^2 \sum_{i=1}^{N} (\phi^a_{i-1} - \phi^a_{i})^2
&\hspace{2cm}&
 \phi_0^a \equiv \phi_N^a  \equiv 0
\eea
 The mass matrix is diagonalized by the orthogonal transformation:
\beq
\phi_j^a = \sqrt{\frac{2}{N}} \sum_{n=1}^{N-1}
\sin \left (\frac{j n \pi}{N} \right )  \phi_{(n)}^a
\eeq
The $n$-th mode  $\phi^a_{(n)}$ has mass equal to $m_n$,
matching at every massive level that of the gauge boson.
Thus, $\phi^a_{(n)}$ corresponds to an adjoint scalar of 5d $\mathcal{N}=2$ gauge multiplets.
\item $N-1$ complex scalars $\phi^s_i = \frac{1}{\sqrt{K}} Tr[\phi_i -v]$,
singlets under the diagonal subgroup. They correspond to flat directions of the
D-term potential and have no counterpart in the 5D supersymmetric $SU(K)$ model.
One can get rid of them~\cite{CSGRKRTE} but
in these paper we kept them in the physical spectrum.
However, if we add a product $U(1)$ hypercharge group to our model
(this implies that the link Higgs also must also have the $U(1)$ charge
$\bf (1/2_i, -1/2_{i+1})$) then the real part of
$\phi^s_i$ turns into a tower of massive scalars with masses
$m_n {\frac{g_0'}{g_0}}$ ($g_0'$ being the gauge coupling of the $U(1)$ interactions),
while the imaginary part is eaten by the massive modes of the
U(1) gauge fields.
\ei

We now turn to the fermionic part of the gauge sector. Similarly as for the link-scalars,
we split the link-Higgsino degrees of freedom into adjoints,
$\psi_{\phi_i}^a = \sqrt{2}Tr[T^a\Psi_{\phi_i}]$,
and  singlets, $\psi_{\phi_i}^s = \frac{1}{\sqrt{K}}Tr[\Psi_{\phi_i}]$,
with respect to the diagonal group.
When the link-Higgs field get vev,
the adjoints $\psi_{\phi_i}^a$ mix with  gauginos.
In addition, we need to add by hand a mass term
$ig_0 v \psi_{\phi_N}^a \chi_N^a$. The full mass matrix becomes:
\beq
\cl = i g_0 v \sum_{i=1}^{N}
(\psi_{\phi_i}^a - \psi_{\phi_{i-1}}^a) \chi_i^a + {\rm h.c}
\eeq
which is diagonalized by:
\bea
&\chi_j^a = \sqrt{\frac{2}{N +1/2}} \sum_{n=1}^{N}
 \cos \left ( \frac{(2j-1)(2n-1)\pi}{2(2N+1)} \right )
  \chi^a_{(n)}&
\nn
&\psi_{\phi_j}^a =  \sqrt{\frac{2}{N+1/2}} \sum_{n=1}^{N}
 \sin \left ( \frac{j(2n-1) \pi}{2N+1} \right ) \psi^a_{(n)}&
\eea
The Dirac fermion
$\lambda_{(n)} = \left ( \begin{array}{c}
\psi_{\phi^a_{(n)}} \\
-\sigma_2 \chi^{a\, *}_{(n)}  \end{array} \right )$
acquires mass:
\beq
\tilde{m}_{(n)} = 2 g_0 v \sin \left ( \frac{(2n - 1) \pi}{2(2N+1)}\right)
 \hspace{2cm} n=1\dots N
\eeq
The singlets $\psi_{\phi_i}^s$ remain massless,
as long as links do not have any $U(1)$ charges.
Apart of the singlets,  in the large $N$ limit
 we recover the mass tower of the BHN model.

The Higgs sector is represented by the set of chiral multiplets
$H_i = (h_i, \psi_{H_i})$ in the fundamental representation of the $i$-th group,
and their antifundamental mirrors  $\tilde H_i = (\tilde h_i, \psi_{\tilde{H}_i})$.
These degrees of freedom acquire their masses from the F-term potential.
In order to obtain the same mass tower as in the gauge sector we choose the superpotential
of the form:
\beq
W = g_0 \left( \sum_{i=1}^{N-1} \tilde{H}_i\Phi_i H_{i+1} -
\sum_{i=1}^{N}v \tilde H_i H_i \right)
\eeq
At the $N$-th site we remove the mirror Higgs
$\tilde h_N$ and  add the scalar mass term $\Delta \cl = g_0^2 v^2 h_N^2$
 (otherwise the Higgs field would not have a zero mode).
After the diagonal symmetry breaking the scalar mass matrix is:
\beq
\cl = -g_0^2 v^2 \sum_{i=1}^{N-1} |h_{i+1}-h_i|^2 -
g_0^2 v^2 \sum_{i=1}^{N}
 |\tilde h_{i-1} - \tilde h_i|^2
 \hspace{1cm} \tilde h_{0} \equiv \tilde h_{N} \equiv 0
\eeq
which is diagonalized by:
 \bea
&h_j =  \sqrt{\frac{2}{N}} \sum_{n=0}^{N-1}
\eta_n \cos \left (\frac{(2j-1) n \pi}{2 N} \right ) h_{(n)} &
\nn
&
\tilde h_j =  \sqrt{\frac{2}{N}} \sum_{n=1}^{N-1}
\sin \left (\frac{j  n \pi}{N} \right ) \tilde{h}_{(n)}
&\eea
Both Higgs towers have masses $m_{(n)}$ but only $h$ has the zero mode.

The mass matrix of the Higgsini is :
\beq
\cl = - g_0 v \sum_{i=1}^{N} \psi_{\tilde{H}_i}
(\psi_{H_{i+1}}-\psi_{H_i}) + {\rm h.c}
\eeq
and is diagonalized by:
\bea
&\psi_{H_j} = \sqrt{\frac{2}{N +1/2}}
 \sum_{n=1}^{N}
 \cos \left ( \frac{(2j-1)(2n-1) \pi}{2(2N+1)} \right ) \psi_{H_{(n)}}&
\nn
&\psi_{\tilde{H}_j} =  \sqrt{\frac{2}{N+1/2}}
\sum_{n=1}^{N}
\sin \left ( \frac{j(2n-1) \pi}{2N+1} \right )
\psi_{\tilde{H}_{(n)}}&
\eea
The Dirac fermion
$\Psi_{H_{(n)}} = \left ( \begin{array}{c} \psi_{H_{(n)}} \\
-i \sigma_2 \psi^*_{\tilde{H}_{(n)}}  \end{array} \right )$
 acquires mass $\tilde{m}_{(n)}$.

As for the quark doublet degrees of freedom,
we again start with a set of fundamental chiral multiplets
$Q_i=(q_i, \psi_{Q_i})$ and their antifundamental mirrors
$\tilde Q_i=(\tilde q_i, \psi_{\tilde{Q}_i})$.
This time we remove the mirror quark
$\psi_{\tilde{Q}_N}$ at the $N$-th site. The squark mass matrix  is:
\beq
\cl = -g_0^2 v^2
\sum_{i=1}^N
 |q_{i+1}-q_i|^2 -
 g_0^2 v^2
 \sum_{i=1}^{N}
 |\tilde q_{i-1} - \tilde q_i|^2
 \hspace{1cm} q_{N+1} \equiv \tilde q_0 \equiv 0
\eeq
which is diagonalized by:
 \bea
&q_j =  \sqrt{\frac{2}{N + 1/2}}
\sum_{n=1}^{N}
 \cos \left (\frac{(2j-1) (2n-1) \pi}{4 N +2 } \right ) q_{(n)}
 &\nn
 &
\tilde q_j =  \sqrt{\frac{2}{N+1/2}}
\sum_{n=1}^{N}
\sin \left ( \frac{j  (2n-1) \pi}{4 N +2 } \right ) \tilde{q}_{(n)}
&\eea
Both squark towers have masses $\tilde{m}_{(n)}$
and there is no zero mode. The quark mass matrix:
\beq
\cl = - g_0 v \sum_{i=1}^{N}
\psi_{\tilde{Q}_i} (\psi_{Q_{i+1}} - \psi_{Q_i}) + {\rm h.c}
\hspace{1cm} \psi_{\tilde{Q}_N} \equiv 0
\eeq
is diagonalized by:
\bea
&\psi_{Q_j} =  \sqrt{\frac{2}{N}}
 \sum_{n=0}^{N-1} \eta_n
 \cos \left (\frac{(2j-1) n \pi}{2 N} \right )
 \psi_{Q_{(n)}}
 &\nn&
\psi_{\tilde{Q}_j} =  \sqrt{\frac{2}{N}}
 \sum_{n=1}^{N-1}
\sin \left ( \frac{j n \pi}{N} \right )
\psi_{\tilde{Q}_{(n)}}
&\eea
The Dirac fermion
 $\Psi_{Q_{(n)}} = \left ( \begin{array}{c} \psi_{Q_{(n)}} \\
-i \sigma_2  \psi^*_{\tilde{Q}_{(n)}}  \end{array} \right )$
 acquires mass ${m}_{(n)}$ and there is single chiral zero mode
 $\psi_Q^{(0)}$.

\setcounter{equation}{0}
\section{Trigonometric sums and determinant}
\label{app:trigo}

We collect in this appendix the results of different trigonometric sums that appear
throughout the paper.
First, the normalizations of the mass eigenstates modes of the different fields follow from
the identities:
\bea
&\ds
\sum_{k=0}^{N-1} \eta_k^2
\cos \frac{(2i-1)k\pi}{2N}
\cos \frac{(2j-1)k\pi}{2N}
=\frac{N}{2}\, \delta_{ij}
\hskip1cm
i,j=1\ldots N
\\
&\ds
\sum_{k=1}^{N-1}
\sin \frac{ik\pi}{N}
\sin \frac{jk\pi}{N}
=\frac{N}{2}\, \delta_{ij}
\hskip1cm
i,j=1\ldots N-1
\\
&\ds
\sum_{k=1}^{N}
\cos \frac{(2i-1)(2k-1)\pi}{4N+2}
\cos \frac{(2j-1)(2k-1)\pi}{4N+2}
=\frac{2N+1}{4}\, \delta_{ij}
\hskip1cm
i,j=1\ldots N
\\
&\ds
\sum_{k=1}^{N}
\sin \frac{i(2k-1)\pi}{2N+1}
\sin \frac{j(2k-1)\pi}{2N+1}
=\frac{2N+1}{4}\, \delta_{ij}
\hskip1cm
i,j=1\ldots N
\eea
The computation of the numerical loop factor in the Yukawa correction to
the Higgs squared mass uses the two relations:
\bea
&\ds
\sum_{k=0}^{N-1} \eta_k^2
\frac{\cos^2 \frac{k\pi}{2N}}{\sinh^2 x + \sin^2 \frac{k\pi}{2N}}
= N \frac{\cosh (2N-1)x}{\sinh 2Nx \, \sinh x}
\\
&\ds
\sum_{k=1}^{N}
\frac{\cos^2 \frac{(2k-1)\pi}{4N+2}}{\sinh^2 x + \sin^2 \frac{(2k-1)\pi}{4N+2}}
= \frac{2N+1}{2} \frac{\sinh 2Nx}{\cosh (2N+1)x \, \sinh x}
\eea
And finally to compute the logarithmic gauge correction
to the Higss squared mass, we need to know:
\bea
&\ds
\sum_{k=1}^{N}
\sin^2 \frac{(2k-1)\pi}{4N+2}
=\frac{2N-1}{4}
\\
&\ds
\sum_{k=1}^{N}
\sin^2 \frac{k\pi}{2N}
=\frac{N-1}{2}
\\
&\ds
\sum_{k=1}^{N}
\cos^2 \frac{(2k-1)\pi}{4N+2}
\sin^2 \frac{(2k-1)\pi}{4N+2}
=\frac{2N+1}{8}
\eea

The computation of the two supertraces (\ref{eq:STrM2})--(\ref{eq:STrM4})
requires the evaluation of the two sums:
\bea
&\ds
\sum_{k=1}^{N-1}
\sin^4 \frac{k\pi}{2N}
=\frac{3N-4}{8}
\\
 &\ds
\sum_{k=1}^{N}
\sin^4 \frac{(2k-1)\pi}{4N+2}
=\frac{6N-5}{16}
\eea

We finish this appendix by a formula for the determinant leading to the mass spectrum of
bosons and fermions after EWSB ($m,p=1\ldots M$, $n,q=1\ldots N$):
\bea
&&
\hskip-1cm
\det
\left(
\begin{array}{cc}
(m_{m}^2-\rho^2) \delta_{mp} +  f_m\, g_p
&
f_m\, c_q
\\
\\
d_m\, g_p
&
(\tilde{m}_{n}^2 - \rho^2)  \delta_{nq}
\end{array}
\right)
\nn
\tv{35}
&&
        \label{eq:det}
=\prod_{m=1}^{M} (m_m^2-\rho^2)
\prod_{n=1}^{N}   (\tilde{m}_n^2-\rho^2)
\left(
1 +
\left( \sum_{m=1}^{M} \frac{f_m g_m}{m_m^2-\rho^2}\right)
\left( 1 - \sum_{n=1}^{N} \frac{c_n d_n}{\tilde{m}_n^2-\rho^2} \right)
\right)
 \eea


\end{document}